# Shortest Path Computation with No Information Leakage

Kyriakos Mouratidis*
School of Information Systems
Singapore Management University
kyriakos@smu.edu.sg

Man Lung Yiu†
Department of Computing
Hong Kong Polytechnic University
csmlyiu@comp.polyu.edu.hk

## ABSTRACT

Shortest path computation is one of the most common queries in location-based services (LBSs). Although particularly useful, such queries raise serious privacy concerns. Exposing to a (potentially untrusted) LBS the client's position and her destination may reveal personal information, such as social habits, health condition, shopping preferences, lifestyle choices, etc. The only existing method for privacy-preserving shortest path computation follows the obfuscation paradigm; it prevents the LBS from inferring the source and destination of the query with a probability higher than a threshold. This implies, however, that the LBS still deduces some information (albeit not exact) about the client's location and her destination. In this paper we aim at strong privacy, where the adversary learns nothing about the shortest path query. We achieve this via established private information retrieval techniques, which we treat as black-box building blocks. Experiments on real, large-scale road networks assess the practicality of our schemes.

## 1. INTRODUCTION

The wide availability of positioning systems and the diffusion of smart-phones has led to an expanding market of location-based services (LBSs). Clients of these services may use their mobile devices to get driving directions to their destination, to retrieve facilities close to their location (e.g., clinics, pharmacies, police stations), to learn who of their social contacts are nearby, etc.

In this paper we consider shortest path queries in transportation networks. Such a network could represent the road segments in a city, where each segment is associated with a cost (e.g., its length or the time required to drive through it). The query computes the sequence of road segments to reach from a source $s$ (usually the client's current location) to a destination $t$ so that the summed cost along the path is minimized. This is one of the most common queries in LBSs. Examples of popular services that support shortest path computation include Google Maps, Map Quest, etc.

*Supported by the Singapore National Research Foundation under its International Research Centre @ Singapore Funding Initiative and administered by the IDM Programme Office.

†Supported by grant PolyU 5333/10E from Hong Kong RGC.

Practical as these services may be, users view them with increasing skepticism. The very nature of the queries may disclose personal information (such as health status, shopping habits, lifestyle choices, etc) which may be tracked and misused by the LBS. Possible forms of misuse include commercial profiling, governmental surveillance, unsolicited and intrusive advertising, etc. The recent example of a leading mobile device company, which had been tracking the locations of its clients without their consent [3, 35], underlines the serious privacy risks in using LBSs. Note that the shortest path query in particular, may disclose information not only about the current position of the client, but also about her intended destination and path taken. The aforementioned risks motivate the development of methods to safeguard client privacy.

The only existing approach for private shortest path queries follows the *location obfuscation* approach [22]. Instead of the query source $s$, this scheme sends to the LBS a set $S$ that includes $s$ and a number of fake source locations. Similarly, it sends to the LBS a set of candidate destinations $T$ that includes $t$ and several fake destinations. The LBS computes the shortest path from every location in $S$ to every location in $T$. Among the $|S| \cdot |T|$ returned paths (where $|S|$ and $|T|$ are the cardinalities of the two sets) the client keeps the one that corresponds to her original source-destination pair. Unfortunately, this approach reveals to the LBS substantial information about the query; e.g., the source $s$ is known to be among the locations in $S$, and $t$ among the $|T|$ candidate destinations.

To avoid such information leakages and provide strong privacy, a promising direction is to apply *private information retrieval* (PIR) [4]. PIR allows a data item (e.g., a disk page) to be retrieved from a server, without the server obtaining any clues about which item was retrieved. Unlike obfuscation, PIR offers cryptographic privacy guarantees, based on reductions to problems that are either computationally infeasible or theoretically impossible to solve. PIR is generally resource-intensive. However, recent PIR protocols achieve practical response times (in the order of seconds over Gigabyte databases [36]), and have been successfully applied to private spatial queries in Euclidean space [28, 19]. As yet, there has been no PIR-based solution for shortest path computation.

Our objective in this paper is to develop practical schemes for answering shortest path queries *without the LBS deducing any information about the queries*. In other words, the LBS knows only that a query is being executed, and can infer nothing else. To meet this requirement, we decide to use existing PIR protocols as building blocks, and rely on their proven security guarantees. Nevertheless, the nature of shortest paths imposes many challenges in developing an efficient PIR-based solution. First, different shortest paths contain different numbers of edges. The result size itself, or the number of data accesses during processing, may reveal information about whether the path is short or long. We must ensure that leakages of this type are prevented. Second, although current PIR protocols achieve reasonable retrieval times, they remain much slower than unsecured disk reads. Therefore, a wise choice of data organization and indexing strategies is of paramount importance to achieve tolerable response times. Third, the PIR building blocks themselves impose limitations. For instance, the PIR protocol we use [36] may only support files up to a certain size, calling for special provisions in solution design. To summarize, our main contributions are:

- We formalize a general methodology that provably achieves total query privacy;
- We develop specific schemes that implement this general methodology;
- We enhance them with novel space optimizations;
- We evaluate our solutions on real road networks and assess their trade-offs.

## 2. RELATED WORK

In this section we survey obfuscation and PIR-based methods for privacy protection in LBSs, along with work on related problems.

### 2.1 Obfuscation Methods

Spatial $k$-anonymity is a type of obfuscation for location privacy that is inspired by the concept of $k$-anonymity in relational databases [33]. The architecture includes (i) the clients, (ii) the LBS that hosts a spatial database and answers queries on it, and (iii) a trusted mediator, commonly referred to as the *Anonymizer*. The clients update the Anonymizer about their most recent locations, and forward to it their queries. Posed a spatial query, the Anonymizer replaces the coordinates of the originating client $u$ with a region (usually a square or a circle) that includes $u$ and at least $k-1$ other clients. This $k$-anonymous region is forwarded to the LBS, which reports back to the Anonymizer possible query answers for any point inside the region. The Anonymizer filters the results, and forwards to $u$ the actual answer to its query. The privacy assurance offered to clients is that, even if the LBS knows the exact locations of all clients, it is unable to identify which among the $k$ clients inside the anonymous region is the query originator.

There exist several spatial $k$-anonymity methods for range and nearest neighbor (NN) queries in Euclidean space [25, 16], as well as adaptations that drop the Anonymizer from the model, and instead have the clients collaboratively form the $k$-anonymous regions [5, 12]. There also exist spatial $k$-anonymity methods for NN processing on road networks [34, 26]; here, instead of a spatial region, a set of road segments is used to anonymize $u$, with the requirement that at least $k-1$ other clients are also located on these segments.

Another class of obfuscation methods use fake locations instead of $k$-anonymous regions. In [8, 20], for instance, the client forwards to the LBS a set of fake query locations along with her actual position. The assumption underlying this technique is that the LBS is unaware of clients' locations, so that it is unable to tell apart the decoys. Another approach is to choose a fake location $u'$ near the client and forward it as the query point [39, 29]. In this setting, nearest neighbor queries can be answered by incrementally fetching from the LBS the NNs of $u'$ and stopping when the set of retrieved data objects is guaranteed to contain the NNs of the actual client's location.

The only method to protect shortest path queries in road networks follows the obfuscation paradigm [22]. This scheme assumes the existence of an *Obfuscator*, which plays a role similar to an Anonymizer, i.e., it serves as a trusted mediator between the clients and the LBS. A client $u$ querying about the shortest path from a source $s$ to a destination $t$, relays its request to the Obfuscator. The Obfuscator appends $s$ and $t$ with a number of decoys, producing obfuscation sets $S$ and $T$, which it then forwards to the LBS. The latter computes all shortest paths from any candidate source in $S$ to any candidate destination in $T$. Upon receipt of these paths, the Obfuscator picks the one that corresponds to the real source and destination, and reports it to the client. To improve performance, [22] suggests that the fake sources and destinations are chosen close to the real $s$ and $t$.

By definition, obfuscation methods disclose some information about the query location, thus providing weak privacy. For instance, spatial $k$-anonymity methods reveal to the LBS that the user lies inside the $k$-anonymous region, which is only a part of the entire data space (and usually a small one). Similarly, in methods that append $u$ with decoys, the LBS is offered a finite set of alternatives for $u$ to be located at. If the LBS can additionally disqualify some decoys (e.g., via contextual knowledge), its chances of guessing correctly the client's location increase. On the other hand, in [39, 29] the LBS does not acquire the actual client location, but it still gains information about her whereabouts, because $u'$ lies in her vicinity. For the specific case of shortest path privacy in [22], the LBS obtains knowledge of a finite set of alternatives for $s$ and $t$ ($|S|$ and $|T|$ candidate locations, respectively) which, moreover, lie near the actual source and destination, providing a rough idea of their positions. Also, this method discloses strong clues about the length and composition of the shortest path itself, since the $|S| \cdot |T|$ paths returned have similar lengths and possibly share many edges too. Our objective in this paper is to prevent leakage of any clues about the query, including any knowledge about $s$, $t$, or the path.

Although not an obfuscation method per se, another approach considered for location privacy is space transformation [18, 37, 38]. In this model, the database owner, who is different from the LBS, maps the data from the original Euclidean space into a transformed space using a keyed function. A querying client $u$ (in possession of the secret key) converts her location into the transformed space and forwards it to the LBS. The latter, although unaware of the secret key and thus unable to map the data and query back to the original space, is still able to compute the query result. [18] supports approximate NN processing, [37] provides exact NN results, and [38] additionally answers range queries. All methods in this category are tailored to spatial queries in the Euclidean space, and are inapplicable to road networks (or graph data in general). Transformation techniques are meant for single client settings, because possession of the secret key by multiple ones implies that any client may collude with the LBS to "decrypt" another's query. Also, transformation schemes are susceptible to access pattern attacks [36]. For example, the LBS may observe the access frequencies of items in the transformed space and use them in tandem with contextual knowledge about the original space to deduce a (partial) mapping between the two spaces.

### 2.2 PIR-based Methods

Private information retrieval (PIR) is a primitive for retrieving data hosted by a server, without the server learning anything about the clients' access patterns [4]. The privacy guarantees of PIR protocols rely on reductions to problems that are either computationally infeasible or theoretically impossible to solve[1]. In this work we use PIR schemes as building blocks in order to exploit these strong guarantees.

Many types of single-server PIR are known to incur prohibitive computation and/or communication overheads for sizable datasets

---

[1] For a complete survey of PIR techniques and an in-depth description of their inner workings, the interested reader is referred to [10].

693

[32]. However, recent *hardware-aided* PIR protocols are shown to be both secure and practical. These protocols utilize a tamper-resistant, *secure co-processor* (SCP) that is installed at the server and is trusted by the clients. For example, [36] features constant communication and amortized polylogarithmic computation cost.

PIR schemes have been applied in the context of spatial queries. The first such method appeared in [11] for NN processing, but relied on a particularly expensive PIR protocol. More recent proposals utilize hardware-aided PIR and report reasonable computation/communication overheads for NN retrieval (a few seconds for Gigabyte databases) [19, 28]. Importantly, [28] asserts that it is not enough to retrieve disk pages from the LBS via a PIR protocol, but the number of pages accessed should be the same for all queries. Otherwise, clues may be given about the data of interest and therefore about the query itself. So far there has been no PIR-based method for shortest path queries.

### 2.3 Relevant Privacy Issues on Graph Data

In this paper we aim at preventing the LBS from deducing clues about the shortest path queries it is called to answer. There exist several streams of work on different (yet related) privacy issues in processing road network data.

One of them focuses on protecting graph data when outsourced to third-party servers or on the cloud (see [9] and references thereof). The main idea in these proposals is to modify the graph so that its exact information is concealed from the hosting server but some of its key topological characteristics are preserved. While this class of methods aims to protect the graph data from the processing server, our objective is to protect the clients' queries; in our model, the road network data are known to the LBS, which may be their owner in the first place.

Another body of related work deals with verifying the accuracy and correctness of the results returned by the processing server [27, 23]. Database authentication schemes have been proposed for shortest path verification in road networks [40]. These approaches are orthogonal to our problem – in our setting the LBS is curious, but not malicious, i.e., it is interested in learning the clients' queries, but would not falsify nor tamper with their results.

Targeting LBS clients that move in road networks, there have also been methods for identity protection (as opposed to location privacy), such as mix-zone techniques [1]. Assuming that the clients need continuous access to an LBS as they walk or drive through a road network, they wish to hide their identity. For this purpose they use pseudonyms to communicate with the LBS. To prevent the linking of pseudonyms with the underlying client identities (through observation of long-term client movements), the pseudonyms change whenever clients enter a mix-zone. Mix-zones are usually placed at road junctions.

Graph problems have also been considered in the semi-trusted model, where multiple parties hold different pieces of information and collaboratively answer a query without disclosing their part of the data to each other. For example, [2] proposes a protocol that allows two entities, which hold different parts of a graph, to compute the all-pair shortest distances in the combined (i.e., complete) graph. In [6] an entity needs to compute a shortest path from a source to a destination without crossing a polygonal obstacle known only to another entity.

## 3. PRELIMINARIES

In this section we frame the problem, define our privacy objective and outline a provably secure methodology. We then present characteristics of SCP technology and of the employed PIR protocol that guide our solution design.

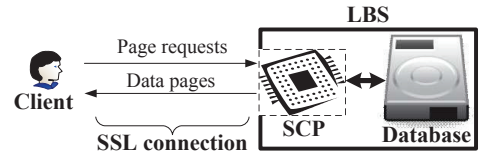

Figure 1: System architecture

### 3.1 Problem Formulation and System Model

**Query:** A road network is modeled as a weighted graph $G = (V, E)$, where $V$ is the set of nodes, and $E$ the set of edges. The nodes $v \in V$ represent junctions, or positions on a road where the traffic conditions or the orientation change, such as road turns. Every edge $e \in E$ connects two nodes and is associated with a positive weight $w(e)$ that models the cost to traverse $e$, e.g., the traveling time from one node to the other, the length of $e$, etc. A path from a source node $s \in V$ to a destination node $t \in V$ is a sequence of edges starting at $s$ and leading to $t$. The cost of a path is defined as the sum of costs across its edges. The path from $s$ to $t$ with the smallest cost is called the *shortest path* and is denoted as $SP(s,t)$. We consider that $E$ includes directed edges and that $s, t$ lie on two network nodes; the discussion easily extends to undirected edges and our contributions apply to query sources/destinations that lie anywhere on the road network (e.g., in between the endpoints of an edge). We assume that all nodes have Euclidean coordinates.

**Architecture:** The road network $G$ is hosted by an LBS – $G$ may be owned by the LBS itself or another entity. The LBS stores on the disk the graph data and any indexing information thereof, organized in equal-sized blocks (pages). The clients of the LBS pose shortest path queries on $G$, and the LBS needs to report the results back to them. A secure co-processor (SCP) is installed at the LBS, and offers a PIR interface for clients to retrieve disk pages from the database of the LBS. Details about the SCP and the PIR protocol employed are given in Section 3.2. Although we assume that the database resides on disk, the PIR interface (and our entire framework) applies to storage in main memory or a solid state drive.

The architecture is visualized in Figure 1. When a client wishes to pose a query, she establishes an encrypted connection (e.g., SSL) with the SCP and answers the query via a multi-round protocol. In each round, the client requests specific disk pages from the SCP, which retrieves them from the database (one by one) in a way oblivious to the LBS. The data fetched determine the page requests in the next round, and so on, until the shortest path is computed.

**Adversary:** The adversary in our model is the LBS. We assume that it knows the client's identity (e.g., via user log-in) or may infer it[2]. The adversary is curious, but not malicious [36], i.e., it wishes to gain information about the clients' queries, yet it executes page access routines correctly, and would not falsify the data in any way. The road network information and its index (if any) are not encrypted, i.e., their plaintext is available to the LBS, who may well be their owner. The adversary is also aware of the processing protocol in use. Its computational power is polynomially bounded (a common assumption that enables the use of cryptographic primitives, such as secure hash functions, etc).

**Security Objective and Privacy Guarantee:** Our objective is to develop practical protocols for processing shortest path queries at the LBS *without the latter deducing any information about the queries*. The database comprises a set of files, e.g., a header file, a graph data file, an index file, etc. Similar to [28], we assert that every shortest path query follows the same *query plan* – this is necessary in order to achieve our privacy goal, as we make clear in the security proof below. Specifically, we ensure that every query (i) executes in the same number of rounds, (ii) in each round it accesses the same files in the same order, and (iii) from each file accessed in a specific round, it retrieves the same number of pages. The query plan is determined by the processing protocol (we will see how) and is publicly available. For example, if the protocol suggests that in the second round 5 pages are fetched from file $F_1$ and then 10 from file $F_2$, every query in its second round must fetch 5 pages from $F_1$ followed by 10 from $F_2$ (in this order). This implies that even though a certain query may need fewer than the specified pages from a file, the protocol pads its requests with dummy page retrievals in order to conform to the query plan. The following theorem proves that our methodology achieves the security objective.

---

[2]Even without user log-in (such as in Google Maps), identification is possible via background knowledge (e.g., user profile/search history), especially if information about the client's source and destination also leaks.

THEOREM 1. *Our methodology leaks no information to the adversary about the shortest path query. Equivalently, every processed query is indistinguishable from any other.*

PROOF. Each page requested from a file is retrieved via an established PIR protocol. Therefore, the adversary is oblivious of which page of the file is being read. What is only visible to the LBS is that a page is being accessed in the specific file. Since all queries follow the same query plan, the number of page retrievals in the various files and their chronological order is identical for all queries, lending the adversary no means to tell any two of them apart. For this reason too, even if the exact same query is re-executed, the LBS is unable to detect that it is processing the same query. Having established that the adversary gains no information from query execution, the proof is completed by the fact that it is also unable to intercept the client's page requests (to the SCP) and the page contents sent back from the SCP (to the client), because they are transmitted via a secure connection (SSL). □

The general methodology described above fulfills our privacy objective. However, the challenge now lies in determining specific processing schemes which (i) ensure that all queries follow the same query plan, and (ii) are practical in terms of performance (e.g., in terms of query response time, space overhead, etc). Before presenting any schemes, we provide some background about hardware-aided PIR that determines our design principles.

## 3.2 Background and Design Considerations

We require a PIR interface that allows clients to securely access the database of the LBS – as explained in Section 2.2, hardware-aided PIR is currently the only practical option. To provide a readily deployable framework, we rely on existing SCP technology and PIR protocols. Hence, we review their properties and limitations.

The SCP is trusted by the clients and installed at the processing server. It has access to the server's disk and may execute a set of cryptographic primitives. SCPs support complete tamper detection, so that clients may remotely assess whether they operate unmolested and unobserved by any potential adversary. The tamper-resistance of SCPs comes at the cost of excessive heat dissipation which, in turn, limits their computation speed and memory capacity. General purpose SCPs are available in the market, such as the IBM 4764 PCI-X Cryptographic Coprocessor.

To fetch disk pages obliviously from the database of the LBS, we employ the protocol of [36] due to its superior performance (note however that alternative PIR protocols could be used). Retrieving a disk page has an amortized computation cost of $O(\log^2 N)$, where $N$ is the total number of pages in the accessed file. The amortized complexity is used because some retrievals may involve reorganization in parts of the file. In absolute terms, a real implementation on IBM 4764 takes around one second to retrieve a page from a Gigabyte file. The communication cost incurred is constant, i.e., the amount of data transferred to the client (via the SSL connection) have the same size as the original disk page read.

The computation cost of the protocol, albeit much smaller than other PIR approaches, is still several times larger than a plain (unsecured) disk read. To ensure the viability of our schemes, *a key objective in our design is to keep the number of pages fetched per query (i.e., per shortest path computation) as small as possible*. This will also limit the communication cost.

Importantly, the protocol of [36] requires that the SCP has at least $c \cdot \sqrt{N}$ memory, where $c$ is a parameter with a typical value of 10. In conjunction with the limited memory on the SCP, this implies that files larger than a certain size cannot be supported – in our experiments, the SCP (IBM 4764) has 32 MByte RAM and may support files up to 2.5 GByte. It is also indicative that the memory capacity in SCP technology increases much slower than, say, hard disk capacity. Therefore, *in our solution design it is essential to keep the database size small*.

A final remark regards our choice to adopt a multi-round methodology, i.e., to have the client lead query processing with repetitive page requests. One could wonder why the processing logic is not completely shipped to the SCP, so that it runs locally the necessary rounds of the protocol, and directly reports to the client the query result (shortest path). The reason is that programming on the SCP is particularly cumbersome, and also that complicated code may lead to prolonged execution due to the aforementioned overheating issues. In our design, we use the SCP merely as an interface to securely fetch specific disk pages (one at a time), using off-the-shelf functionality in order to ensure direct applicability.

## 4. BASELINE SOLUTIONS

There exist several approaches to process shortest path queries in the literature, but they are unsuitable to our model. The straightforward way to answer the query is to invoke a disk-based version of Dijkstra's algorithm [7], A* search [14], etc. Data organization aside, performance in our setting would be prohibitive. As elaborated previously, *all queries in a secure scheme must perform the same number of page retrievals*. This implies that every query will incur the same processing cost as the costliest possible shortest path computation. It is a known fact that for certain source-destination pairs (e.g., the anti-diagonal nodes in $G$), these algorithms access almost the entire road network. This would bound every query to incur a cost equivalent to accessing all pages in the database.

There are indexing methods where search runs first on a hyper-graph in order to guide processing in the underlying network $G$, such as HiTi [15]. These approaches rely on the same expansion principles as Dijkstra's algorithm in the voluminous (due to heavy materialization) hyper-graph, and run into similar problems.

Several pre-computation methods also exist, such as *Landmark* [13], *Arc-flag* [21], *SPQ* [31], etc. The idea is to materialize some information together with each node or edge, so that when the node/edge is reached by network expansion, it helps narrow down the possible hops to adjacent nodes. Adapting these methods to our setting reduces in part the deficiencies of plain network search. However, performance remains problematic because distant source-destination pairs still require reading a large portion of the database (thus bounding any query to an equally high processing cost). For the sake of comparison with our advanced schemes,

695

we present the two best-performing adaptations we constructed from this category, based on Landmark and Arc-flag.

**Adaptation of Landmark (LM):** Landmark [13] chooses a number of anchor nodes in $G$ and pre-computes for each $v \in V$ the shortest path costs (from $v$) to the anchors. The vector of costs, called Landmark vector, is kept with $v$ and helps compute estimates for the cost of $SP(v, t)$ to the destination $t$. This information is utilized by A* search which visits first nodes with a small estimated cost to $t$. We adapt Landmark to our setting and use it as a baseline in our experiments; we refer to this scheme as LM.

To enhance performance, we exploit locality. In particular, we first partition $G$ into regions. For each region we allocate one disk page and inside store the information (i.e., the adjacency lists and Landmark vectors) of all its nodes. The resulting file is denoted as $F_d$. It is essential that the partitioning method does not waste space, i.e., it leaves in every disk page as little empty space as possible, in order to keep the database small. To produce regions that fulfill this requirement we use a method described in Section 5.6.

In the first round of processing, the querying client requests for and receives a *header file*, denoted as $F_h$. This file includes the partitioning information, i.e., it allows mapping any point that lies on the network to the region that contains it. Additionally, for each region it indicates the page number in $F_d$ that holds its data. Finally, it also specifies the query plan (the plan's derivation is described after the algorithm outline). The header is small and must be accessed by any querying client. It is therefore retrieved from the LBS in its entirety, without using the PIR interface.

Having received the header, the client locally maps its source and destination into the containing regions $R_s$ and $R_t$, termed *the source and the destination region*, respectively. In round two, she fetches from $F_d$ the pages that hold the data of these two regions (via the SCP and the secure connection, so that the LBS is oblivious of which regions these are). She initializes an A* search at $s$ using the $R_s$ data. When the search encounters a node that belongs to another region, a new round of processing is initiated and the corresponding $F_d$ page is fetched via the PIR interface, and so on, until the destination $t$ is reached. Note that all queries must abide by the query plan, which means that upon reaching $t$, the client may need to make dummy requests until the necessary number of page retrievals is reached.

To determine the query plan, we execute the algorithm (without dummy requests) from all possible sources $s \in V$ to all possible destinations $t \in V$, and record the maximum number of pages needed from $F_d$. Observe that in LM the query plan is simply defined by the number of pages retrieved, because every round fetches exactly one page from $F_d$, with the exception of the first round, which fetches two (for $R_s$ and $R_t$).

**Adaptation of Arc-flag (AF):** Arc-flag [21] requires partitioning the road network into regions. For each edge $e \in E$, it keeps a bit-vector where every bit corresponds to a region – the bit for a region is set to 1 only if there is a shortest path from one endpoint of $e$ to a node in that region that passes through $e$. With this information, processing a shortest path query only considers edges whose bit for the destination region is 1. We construct a second baseline, termed AF, that relies on Arc-flag.

The adaptation is similar to LM. A major difference, however, is that we drop the requirement that the data of a region (adjacency lists of nodes and bit-vectors of edges) must fit in a disk page. For large networks this would lead to numerous regions, and thus huge bit-vectors. Instead, we allocate for each region a fixed number of pages, to be retrieved together during query processing. The number of pages per region is a parameter of the method.

To conclude the discussion about pre-computation techniques, we stress that full materialization is also inadequate for our setting. This approach would compute and materialize the shortest paths for every possible source-destination pair, so that the result for any query could be looked up directly. The problem is that, even for small road networks, the space needed to store all paths is orders of magnitude larger than the maximum supported by the PIR interface. For the smallest network in our evaluation (Oldenburgh, with around 6K nodes), this approach requires approximately 20 GByte, which increases cubicly with the network size.

The spatial network literature includes methods for shortest path computation on the air [17], i.e., where the road network data are periodically broadcast, and the clients tune in the channel to process their queries locally. Their objective is to construct a broadcast cycle and inside distribute indexing information in order to minimize (i) the time that the client keeps its receiver on and (ii) the distance (in the broadcast cycle) between the first and the last data packet needed for query processing. These methods are inapplicable to our case due to the different nature of the problem (e.g., on the air there is no random access because the clients cannot control the broadcast schedule – once missed, a packet can only be received in the next broadcast cycle). However, the technique in [17] includes the idea of partitioning the network and broadcasting, for every pair of source-destination regions, the intermediate regions that may appear in a path between them. The pre-computation in Section 5.2 employs a similar idea.

## 5. CONCISE INDEX SCHEME

The main performance factors in our design, as established in Section 3.2, are query processing cost and database size. The former is linked directly to the maximum number of pages needed for any possible source-destination pair (due to the fixed query plan requirement). The latter, i.e., database size, indirectly affects the retrieval cost (recall that the time to fetch a page via the SCP increases polylogarithmically with the number of pages in the file), but the primary reason to keep it small is because the PIR interface may support files up to a certain size only.

Our first scheme is termed *Concise Index* (CI). It features a minimal space overhead and a manageable query processing cost. In CI the database consists of four files, namely the *header*, the *look-up*, the *network index* and the *region data* file; we denote them as $F_h$, $F_l$, $F_i$, $F_d$, respectively. Their roles are as follows.

- *Header*: CI partitions the network into regions. The header helps the client map her source and destination to their host regions. It also includes the query plan.

- *Look-up*: It enables browsing the network index file.

- *Network index*: It includes pre-computed information that helps guide the shortest path search.

- *Region data*: It stores the actual network information of each region, i.e., node coordinates, adjacency lists, etc.

We first present the pre-processing steps in CI, i.e., network partitioning and pre-computation (Sections 5.1 and 5.2). Next, we describe the exact contents of each file (Section 5.3). Then, we discuss the derivation of the query plan and the query processing algorithm (Section 5.4). Finally, we propose space optimization techniques (Sections 5.5 and 5.6).

### 5.1 Network Partitioning

CI, as well as subsequent schemes, relies on a partitioning of the road network into regions. The choice of partitioning method is important. One requirement is that it must be easily representable

696

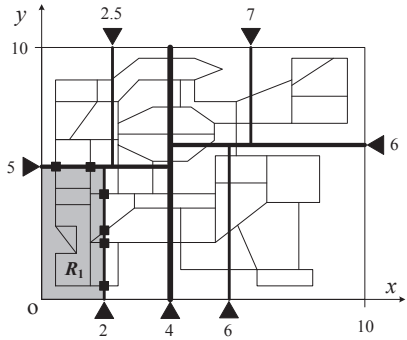

Figure 2: KD-tree partitioning and border nodes

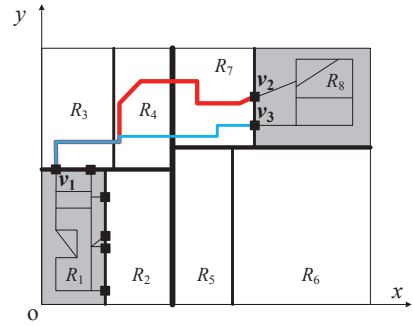

Figure 3: Shortest paths between border nodes

in terms of Euclidean coordinates. The reason is that clients are unaware of node or region identifiers[3], and may only express their source and destination in terms of Euclidean coordinates. Another requirement is that space is not wasted. Since regions are to be placed on disk and it is essential to keep the database small, we need to leave as little unutilized space in each page as possible. Importantly, the partitioning should facilitate query processing, implying that regions should be chosen such that shortest paths are likely to cross as few of them as possible. Last but not least, the partitioning information should be expressible in a concise form, because it will be sent to the clients (as part of $F_h$) over a communication network.

A simple partitioning method is to superimpose a KD-tree (in Euclidean space) on the road network. This technique produces regions of comparable quality (in terms of facilitating shortest path computation) to more sophisticated and complex alternatives [24]. Additionally, the tree structure (which essentially determines the mapping between Euclidean coordinates and network regions) can be represented in a very concise form.

Each leaf of the tree holds the nodes that lie inside its spatial extent; a node's information includes its identifier, its coordinates and its adjacency list (i.e., the list of adjacent nodes and the weights of the corresponding edges). Every leaf determines a region and is associated with a region identifier $R_i$. Figure 2 illustrates the KD-tree partitioning of a sample road network. The bold lines correspond to the split lines of the tree nodes. Region $R_1$ is defined by the leaf shown shaded, and holds the information of all nodes inside. The tree structure can be represented simply by the splitting coordinate (either on the $x$ or $y$ axis) used in every node of the tree, e.g., the first split is at $x = 4$, followed by splits at $y = 5$ and $y = 6$ in the left and right child of the root, and so on.

KD-tree partitioning fulfills all requirements we set, except for high disk page utilization. The idea in CI is that node information for each region is placed in a single disk page[4]. That can be enforced by splitting the tree nodes until the network information in every leaf fits in a page. The problem, however, is that this may leave up to 50% of the page empty, leading to an unnecessarily large database. We developed a KD-tree construction method that minimizes unutilized space and effectively reduces the database size. We leave the details of this technique for Section 5.6.

---

[3]Node and region identifiers are a matter of naming during database creation, and cannot be assumed known to the client in advance.

[4]Note that placing region information contiguously on the disk (i.e., ignoring physical page boundaries) leads to regions that may cross over to a second page. We wish to avoid this for performance reasons that will become clear shortly.

## 5.2 Pre-computation

CI pre-computes and materializes some shortest path information. Key in this process is the notion of *border nodes*. These are intersection points of the network edges with the splitting lines of the KD-tree. In Figure 2, for example, region $R_1$ has 6 border nodes, represented as solid squares. Border nodes are treated as normal network nodes during pre-processing, but they are discarded afterwards (i.e., not stored in any file).

The fundamental property of border nodes is that any path starting from a source $s$ inside some region $R_s$ to a destination outside of it must pass through one of the border nodes of $R_s$. Similarly, any path to a destination $t$ in region $R_t$ (from a source outside of it) passes through a border node of $R_t$. Consider a shortest path $SP(s, t)$ and let $v$ and $v'$ be the border nodes of the source and destination region, respectively, that appear in this path. Due to its cost minimality, $SP(s, t)$ is guaranteed to include $SP(v, v')$. The above facts combined suggest that $SP(s, t)$ passes necessarily via $SP(v, v')$ for some border node pair $(v, v')$. In Figure 3, assume that $s$ is somewhere in $R_1$ and $t$ in $R_8$. If the shortest path $SP(s, t)$ passes through border node $v_1$, it necessarily includes either $SP(v_1, v_2)$ (shown red) or $SP(v_1, v_3)$ (shown blue), where $v_2$ and $v_3$ are the border nodes of $R_8$.

Based on this observation, CI computes for every pair of regions $R_i, R_j$ the shortest paths from all border nodes in $R_i$ to all border nodes in $R_j$. Let $S_{i,j}$ be the set of intermediate regions crossed by at least one of these paths. For example, the consideration of border node pair $(v_1, v_2)$ in Figure 3 would include (the identifiers of) $R_3, R_4, R_7$ into region set $S_{1,8}$. By definition, any shortest path from a source in $R_i$ to a destination in $R_j$ may pass only through $R_i$, $R_j$ and regions in $S_{i,j}$. This pre-computation process is also necessary for $S_{i,j}$ sets where $i = j$ (i.e., when source and destination regions are the same) because a shortest path between border nodes of $R_i$ might still pass through a neighboring region.

## 5.3 File Formation

After partitioning and pre-computation, the CI files are formed.

**Region Data File ($F_d$):** As mentioned previously, $F_d$ includes exactly one page for every region $R_i$. Inside it keeps the network information of $R_i$, including node identifiers, their adjacency lists and incident edge weights.

**Network Index File ($F_i$):** $F_i$ contains the pre-computed $S_{i,j}$ information. The region sets $S_{i,j}$ are stored into pages in ascending order of composite key $(i, j)$. They are placed contiguously into pages, with the objective of minimizing the total number of pages each of them spans. In particular, for $S_{i,j}$ sets with size smaller than a page (as in the vast majority of cases), we prevent them from stretching over two pages. This implies that during file formation, if the free space in a page is not enough to host the next $S_{i,j}$ set (in $(i, j)$ order), the space is left unutilized and the region set is placed in the next page of the file. This is important in order to reduce the PIR retrieval cost per region set. The same reasoning applies to (the rare case of) region sets larger than a page; an $S_{i,j}$ set with size, say, 2.2 times the page capacity, is not allowed to stretch over four pages – if necessary, its data start in a new page in order to span three pages in total.

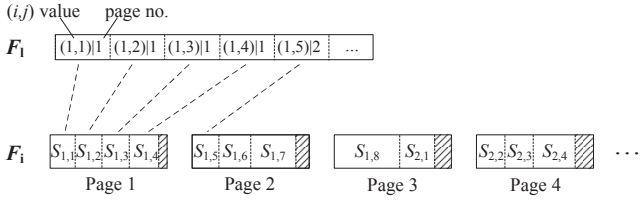

**Figure 4:** $F_l$ **and** $F_i$ **example**

Figure 4 (in its lower part) illustrates an example of $F_i$. The striped space at the end of the pages is unutilized. The figure demonstrates the general case, where edges are directed. In case of an undirected graph, sets $S_{i,j}$ and $S_{j,i}$ would be identical, and hence region sets $S_{i,j}$ where $i > j$ would be omitted from the network index file.

**Look-up File** ($F_l$)**:** $F_l$ is essentially a dense index over $F_i$, as shown in Figure 4. Specifically, for every $(i, j)$ pair, $F_l$ stores a look-up entry that indicates the page number in $F_i$ that holds region set $S_{i,j}$. The $F_l$ entries are sorted on composite key $(i, j)$. The pages in $F_l$ are packed, i.e., each stores the maximum possible number of look-up entries. This implies that for any pair $(i, j)$, a division by that number indicates the $F_l$ page that holds the corresponding look-up entry (which in turn leads to the actual $S_{i,j}$ data in $F_i$). Note that $F_l$ is much smaller than $F_i$ because a look-up entry takes up less space than the average $S_{i,j}$ set. Also, its size can be reduced by omitting the $(i, j)$ values because they are implicitly defined by their order in $F_l$.

**Header File** ($F_h$)**:** The header includes the KD-tree information that allows mapping $s$ and $t$ to their host regions. For each leaf of the KD-tree (i.e., for each region) the header also stores (i) a region identifier (e.g., $R_1$, $R_2$, etc), and (ii) the page number in $F_d$ that holds the actual network information of the region. The header additionally specifies the query plan and meta-data about the other three files (e.g., filename, size, record length for $F_l$, etc). $F_h$ is small and needs to be downloaded by every client who wishes to pose a query. Therefore, it discloses no information about the query itself, and is downloaded in full directly from the LBS, without involving the PIR interface.

## 5.4 Query Processing

Consider a client who wishes to know the shortest path from source $s$ to destination $t$, and ignore the query plan for the time. In the first round of processing, the client receives the header file $F_h$. Based on the coordinates of $s$ and $t$, she uses the KD-tree information to determine the source and destination regions $R_s, R_t$. Note that there is no requirement that $s$ and $t$ are network nodes; they could lie anywhere on the road network.

In the second round, the client uses the PIR interface and fetches the page in the look-up file $F_l$ that corresponds to pair $(s, t)$. She extracts the look-up entry for the specific pair and learns the page number in the network index file $F_i$ that stores the $S_{s,t}$ set. In the third round, she fetches that page from $F_i$ (via the SCP); if $S_{s,t}$ stretches to nearby pages, they are also retrieved (this incorporates an implementation detail elaborated shortly).

In the fourth round, the client requests (via the SCP) the pages of $F_d$ that include the network information of $R_s$, $R_t$, and all regions in $S_{s,t}$. Upon receipt of these data, she possesses a subgraph of $G$ that is guaranteed to contain the desired shortest path. $SP(s, t)$ is computed using Dijkstra's algorithm in this subgraph.

**Query Plan:** In addition to the above accesses, the query plan may require extra (dummy) page retrievals. In the first round the entire $F_h$ is downloaded. In round two, there is always a single page fetched from $F_l$. In round three, we force each query to retrieve as many pages from $F_i$ as the maximum number of pages spanned by any $S_{i,j}$ set. This means that even if a single $S_{i,j}$ set spreads over three pages in the file (while every other fits in one or two), any query will need to make three retrievals in $F_i$. An important implementation detail here is that the client does not know in advance how many pages $S_{s,t}$ spans, but she knows from the query plan that the maximum it could be is three. Therefore, it requests for the page indicated by the look-up entry for pair $(s, t)$ plus the subsequent two pages[5].

Regarding round four, let $m$ be the maximum number of regions inside any $S_{i,j}$ set. Recall that each region's data fit in a single page of $F_d$. The query plan ensures that every query accesses $m + 2$ pages in $F_d$ (the extra two pages account for $R_s$ and $R_t$).

## 5.5 Index Compression

In this section we describe a technique to bring down the space overhead of CI (and, indirectly, the PIR retrieval cost too). Specifically, we reduce the size of the network index file. The crucial observation is that the $S_{i,j}$ sets for nearby $(i, j)$ pairs have significant overlaps due to locality. This motivates an in-page compression mechanism, which takes place as $F_i$ is being formed.

When storing an $S_{i,j}$ set into a disk page, we check which of the region sets already in this page has the largest overlap (i.e., the largest number of common elements) with $S_{i,j}$. Let this set be $S_{k,l}$. We use $S_{k,l}$ as reference for $S_{i,j}$ and only store its *delta information*, determined as follows. Let $n$ be the number of elements (region identifiers) that appear in $S_{i,j}$ but not in $S_{k,l}$. These region identifiers must be definitely indicated in the delta information to ensure correctness of shortest path computation.

If the number of elements in $S_{k,l}$ plus $n$ is no larger than value $m$ (introduced in Section 5.4), the delta includes just these $n$ region identifiers. Expressing the $S_{i,j}$ information this way, implicitly inflates its contents (by those region identifiers that belong to $S_{k,l}$ but not to $S_{i,j}$). This, however, incurs no response slow-down, because anyway a shortest path query from $R_i$ to $R_j$ needs to perform a number of dummy requests in the region data file in order to reach $m + 2$ in total (so as to adhere to the query plan) – essentially, our compression strategy replaces some of the dummy requests with fetching unneeded region data.

On the other hand, if the number of elements in $S_{k,l}$ plus $n$ exceeds $m$, the delta information of $S_{i,j}$ must additionally specify region identifiers *not to be fetched*. These identifiers should belong to $S_{k,l}$ but not to $S_{i,j}$. We need to indicate as many such elements for exclusion, as to inflate $S_{i,j}$ only up to a cardinality of $m$. For example, assume that $S_{i,j}$ includes elements $R_1, R_2, R_8$, while $S_{k,l}$ contains $R_1, R_2, R_3, R_4, R_5$. The delta information of $S_{i,j}$ definitely includes $R_8$. If $m \geq 6$, there is no need to indicate exclusions. A shortest path query from $R_i$ to $R_j$ will fetch from $F_d$ all 5 regions in $S_{k,l}$ plus $R_8$ (plus a number of dummies, if necessary). On the other hand, if $m = 5$, the delta information must exclude at least one of $R_3, R_4, R_5$.

---

[5] As boundary-case exception, if the look-up entry leads to either of the last two pages in $F_i$, the client requests for its last three pages.

Note that our $F_i$ compression mechanism is applied within individual pages and not across them. That is, we do not use reference sets $S_{k,l}$ outside the page meant to hold $S_{i,j}$ because that would imply additional page requests in $F_i$ and increase the response time.

## 5.6 Packed Partitioning

As explained in Section 5.1, we choose a KD-tree to perform network partitioning for a variety of reasons. However, the standard KD-tree would leave up to 50% unutilized space in $F_d$. Here we propose a tree packing mechanism that guarantees high space utilization (over 95% in our experiments). We need to allocate exactly one page for each tree leaf (region), and inside store the information of all nodes that lie in its spatial extent (including their coordinates, adjacency list, etc). The difficulty of the problem lies in the variable length of node information, because adjacency lists of different nodes have different sizes. Assuming that the largest node information takes up $z$ bytes, our strategy guarantees that all pages in $F_d$ (but the last) have no more than $z$ unutilized bytes. To achieve this, we construct an unbalanced KD-tree.

Suppose that each disk page has size $B$ bytes, and assume that the first split (at the root of the KD-tree) is meant to be vertical, like in Figure 2. We sort the information of all nodes based on their $x$ coordinate and place them contiguously to form a byte-stream. The split is made at the $(2^i \cdot (B - z))$-th byte of the stream, where $i$ is the smallest integer for which the split position is to the right of the middle byte in the stream. All nodes to the left of the split position (including the node that owns the information stored in the specific position) are placed into the left child of the root.

Consider now the subtree at the root's left child. The child is split iteratively in a way similar to a plain KD-tee, i.e., splits happen at the middle byte of the node information stream. The power of the algorithm is that we can *arbitrarily* push the node that overlaps with the split position either to the left or right of the split, without fear of eventually overflowing any of the resulting leaves (pages). This is the case, because when we had split the root we implicitly allowed a leeway of $z$ bytes per region (which is enough by definition to fit any node's information). On the other hand, the splits lead eventually to $2^i$ leaves. Having squeezed at least $2^i \cdot (B - z)$ bytes into $2^i$ leaves, we are sure to utilize at least $B - z$ bytes per page.

The process applied to the root is now recursively repeated for its right child. That is, we sort its contents on $y$ coordinate. In the resulting byte-stream, we perform a split at position $2^j \cdot (B-z)$ for the smallest integer $j$ that puts the split to the right of the middle byte. Splits in its own left child happen simply at the middle byte, leading to $2^j$ leaves that hold at least $2^j \cdot (B - z)$ bytes of data. In its right child, a byte-stream is formed and split in a way similar to the root, and so on.

## 6. PASSAGE INDEX SCHEME

As we show in the experiments, CI requires little extra space compared to simply storing the raw network data. However, the longest paths in $G$ may span a considerable number of regions (implying that value $m$, in Section 5.4, may be large). That leads to a significant number of PIR accesses in $F_d$ which dominate the response time. Motivated by this fact, we propose the *Passage Index* (PI) scheme – with the use of more space, PI achieves a drastic reduction in the number of pages needed, and thus in response time.

In PI, instead of having the client retrieve all intermediate regions between $R_s$ and $R_t$, we materialize an exact subgraph that links them. Specifically, pre-computation is the same as in CI. However, instead of keeping $S_{i,j}$, we record for every pair of $R_i$ and $R_j$ the exact edges that appear in one or more shortest paths between their border nodes. Essentially, these edges define a subgraph $G_{i,j}$, such that every shortest path from $R_i$ to $R_j$ is guaranteed to pass entirely through the union of $R_i$, $R_j$ and $G_{i,j}$. In the example of Figure 3, $G_{1,8}$ includes, among others, the edges that belong to the two shortest paths (shown in red and blue).

PI involves four files, formed as explained in Section 5.3, the difference being that the network index file includes the $G_{i,j}$ information (instead of $S_{i,j}$). Placement into physical pages follows the same principles. In query processing, however, there are only three rounds. The first two are identical to CI, while the third fetches (i) from $F_i$ the subgraph $G_{s,t}$ that corresponds to the source and destination regions, and (ii) the two pages in $F_d$ that hold the network information of $R_s$ and $R_t$. Regarding the query plan, let $h$ be the maximum number of pages spanned by any subgraph $G_{i,j}$ in the network index file. Each shortest path computation should retrieve in the first round the entire header (directly from the LBS, without involving the SCP), in the second round one page from $F_l$, while in the third round exactly $h$ pages from $F_i$ and two pages from $F_d$. Note that if $h = 1$, which may be the case for a small network, PI answers the query with only four PIR accesses.

In PI the network index file vastly dominates the space requirements. To reduce its size, we observe that subgraphs $G_{i,j}$ exhibit locality (i.e., for nearby $(i, j)$ pairs the subgraphs share many common edges) and apply a similar compression to Section 5.5. When inserting $G_{i,j}$ into an $F_i$ page, we chose as reference the subgraph in the same page with the largest number of common edges. The delta information in this case does not need to indicate exclusions (as they do not affect the query plan), but simply specifies the edges in $G_{i,j}$ that are missing from the reference subgraph.

As we show in Section 7, the compression strategy reduces space drastically. However, the network index file may still be voluminous. In our experiments, the database size in PI is two orders of magnitude larger than CI, which renders it inapplicable to the largest networks we used (it exceeds the maximum file size supported by the PIR interface). Besides PIR-imposed limitations, the LBS may need to host multiple databases and therefore allocate only a specific amount of space to support queries on $G$. Below we propose two techniques that sacrifice in part the efficiency of PI (i.e., response time) in order to reduce space requirements.

**Hybrid Scheme (HY):** PI performs a small number of page retrievals, but needs a lot of space. On the other hand, CI requires little space, but if value $m$ (i.e., the maximum number of regions across all $S_{i,j}$ sets) is large, response time will suffer due to the PIR access cost. HY is a hybrid between the two schemes that features faster response time than CI and smaller space requirements than PI. Importantly, the space-time trade-off may be tuned to suit the application requirements.

Using the same pre-computation as CI, we form region sets $S_{i,j}$ for all $(i, j)$ pairs. Instead of placing them directly into $F_i$, we iteratively identify the region set with the largest number of elements (i.e., the one that determines value $m$, and therefore the number of PIR accesses) and replace it with its $G_{i,j}$ counterpart. This replacement procedure increases gradually the size of $F_i$, and reduces the number of PIR accesses, i.e., accelerates processing. Replacement stops when a desirable trade-off between space and response time is struck, or when $F_i$ reaches the maximum permissible size. The network index file stores for every $(i, j)$ pair either a region set $S_{i,j}$ or a subgraph $G_{i,j}$.

The header is similar to CI/PI. In $F_l$, all look-up entries have the same format, regardless of whether they lead to a region set or subgraph in $F_i$, and are intermixed transparently in the file. A key difference from previous schemes is that $F_i$ and $F_d$ are concatenated into a single file. The reason will become clear shortly.

Assume for simplicity that each of the $S_{i,j}$ sets (that were not replaced by subgraphs during pre-processing) fits in one page of $F_i$. Query processing is identical to CI/PI in the first two rounds (accessing $F_h$ and $F_l$). In the third round, after retrieving the lookup entry for pair $(s,t)$, we access $F_i$ and fetch $S_{s,t}$ or the first page of $G_{s,t}$ (whichever was stored in $F_i$ during pre-processing). If it is $S_{s,t}$, in the fourth round we access from $F_d$ the pages that store $R_s, R_t$ and all regions in $S_{i,j}$. If it is $G_{s,t}$, the fourth round fetches from $F_i$ the subsequent pages of the subgraph information (recall that the third round only fetched the first page of $G_{s,t}$), and from $F_d$ the network data of $R_s, R_t$.

Note that if $F_i$ and $F_d$ are separate files, the adversary can observe how many pages are accessed from each, and infer whether the client's query was answered via a region set or a subgraph. This is a leakage we cannot afford; if the query was answered via a region set $S_{i,j}$, the adversary (which is aware of the replacement process and its parameters) can immediately narrow down the possible source-destination regions to those who were not chosen for replacement during pre-computation, and vice versa. Hence, $F_i$ and $F_d$ are concatenated into the same physical file. The query plan requires accessing one page from this combined file in round three. For round four, we compute the maximum number of pages needed for any $(i,j)$ pair and assert that every query fetches as many.

A final remark about HY regards the general case where not all $S_{i,j}$ sets fit in a single page. Let $r$ be the maximum number of pages that an (un-replaced) region set spans in $F_i$. In this case, any query in round three must access exactly $r$ pages from the network index file, be it for a region set or a subgraph. These $r$ pages are consecutive in $F_i$ to guarantee that if the $(s,t)$ pair corresponds to a region set, $S_{s,t}$ will be read in full. If it is a subgraph (i.e., $G_{s,t}$), its remaining pages (past the $r$-th) will be read in round four.

**Clustered PI (PI*):** Another alternative to reduce the space requirements of PI is to keep the scheme as is, but allocate more than one page per region. We name this variant *clustered PI* and denote it as PI*. Partitioning uses the packed KD-tree approach in Section 5.6, with the extended leaf/region capacity. The exact number of pages per region is a system parameter that determines the trade-off between space and time.

The above amendment effectively reduces the size of $F_i$ because it leads to (i) fewer regions, and (ii) fewer border nodes in total (because there are fewer KD-tree splits). In turn, fewer regions imply fewer subgraphs $G_{i,j}$ in $F_i$. Also, fewer border nodes imply materializing shortest path information (in the form of subgraphs) for fewer border node pairs. The more pages allocated per region, the smaller the network index file. Size aside, the construction and usage of $F_i$ is similar to the original PI.

Regarding $F_d$ formation, the pages that correspond to a region are placed contiguously on disk. As per normal, $F_d$ is accessed in the third round of processing to retrieve the network data of $R_s$ and $R_t$. However, all pages of these two regions must be fetched. This is the only difference from the original PI processing, which increases (to a controllable degree) the response time. For instance, if 3 pages are allocated per region, the clustered PI retrieves 6 pages of $F_d$ per query (instead of just two in the original scheme).

## 7. EXPERIMENTS

In this section we evaluate empirically our schemes on real road networks. We also quantify the effectiveness of individual optimizations.

**Table 1: Road networks**

| Road network | Number of nodes | Number of edges |
|---|---|---|
| Oldenburg (Old.) | 6,105 | 7,029 |
| Germany (Ger.) | 28,867 | 30,429 |
| Argentina (Arg.) | 85,287 | 88,357 |
| Denmark (Den.) | 136,377 | 143,612 |
| India (Ind.) | 149,566 | 155,483 |
| North America (Nor.) | 175,813 | 179,179 |

**Table 2: System specifications**

| System parameter | Value |
|---|---|
| Disk page size | 4 KByte |
| Disk seek time | 11 ms |
| Disk read/write rate | 125 MByte/s |
| SCP read/write rate | 80 MByte/s |
| SCP encryption/decryption rate | 10 MByte/s |
| Communication bandwidth | 384 Kbit/s (48 Kbyte/s) |
| Communication round-trip time | 700ms |

### 7.1 Experiment Setup

The evaluation considers our advanced PIR-based methods (CI, PI, HY, PI*) and the baseline competitors (LM, AF; described in Section 4). CI and PI are parameterless, whereas HY and PI* involve a parameter each to tune their index size. By default, the methods incorporate all optimizations presented in the paper (e.g., the packed partitioning in Section 5.6, the index compression in Section 5.5, etc). We also provide measurements for an obfuscation method based on [22] (OBF); this is for the sake of a performance indication only, because privacy-wise OBF leaks substantial information about the client queries, as explained in Section 2.1.

We implemented all methods in C++ and conducted experiments on a machine with an Intel Core2 Quad CPU 2.83 GHz and 4 GByte of RAM. Table 1 describes the road networks used in our evaluation: Oldenburg was obtained from Brinkhoff et al.[6] and the rest from the Digital Chart of the World[7]. Our machine uses a Seagate 320 GB (7,200 RPM) hard disk,[8] with 11 ms disk seek time, 125 MByte/s disk read/write rate, and 4 KByte disk page size. Similar to [36], we adopt the *IBM 4764 PCI-X Cryptographic Coprocessor*[9] as the SCP and strictly simulate its performance. The SCP has 32 MByte memory and may support file sizes up to 2.5 GByte. Table 2 summarizes the specifications of the SCP and the hard disk (these values determine the time to retrieve a disk page via the PIR interface, as detailed in [36]). The client communicates with the LBS using a link with round trip time of 700ms and bandwidth 384 Kbit/s (i.e., 48 Kbyte/s) – this corresponds to a moving client connected via a 3G network [30].

The average *response time* of a method is measured by running a workload of 1,000 shortest path queries. It denotes the elapsed time from query submission until obtaining the shortest path result. It consists of: (i) server processing time, (ii) communication time, and (iii) client-side computation time. For the obfuscation method (OBF), component (i) refers to the processing of obfuscated queries at the server. For the PIR-based methods, component (i) corresponds to the PIR time for fetching disk pages from the database.

---

[6]http://iapg.jade-hs.de/personen/brinkhoff/generator/
[7]http://www.maproom.psu.edu/dcw/
[8]http://www.seagate.com/www/en-us/products/desktops/barracuda hard drives/
[9]http://www-03.ibm.com/security/cryptocards/pcixcc/overhardware.shtml

## 7.2 Tuning of Baseline Schemes

The performance of baseline approaches is affected by their parameter values, namely, the number of anchor nodes in LM, and the number of regions (i.e., the number of Arc-flag bits kept with each edge) in AF. For fairness, we tune their parameters so that they achieve the shortest possible response time. Since the optimal values vary for different road networks, we fine-tune LM and AF individually for each dataset.

In Figure 5 we describe LM tuning for the Argentina network. The plots show its response time and space requirements with respect to the number of anchor nodes (i.e., the length of the Landmark vector kept with each node). LM achieves the shortest response time when 5 anchors are used. Figure 5(a) shows that too few anchors lead to slow execution. The reason is that the Landmark vectors are not descriptive enough to effectively narrow down the search space (i.e., to guide the A* search), which results in fetching too many pages. On the other hand, the more anchors used, the larger the $F_d$ file (see Figure 5(b)). Given that PIR accesses become more expensive when the file size increases (as explained in Section 3.2), using too long Landmark vectors harms responsiveness.

The fine-tuning (and the trends) of AF is similar. For Argentina network, AF achieves the best response time when the number of Arc-flag bits is 8. We omit the charts for brevity.

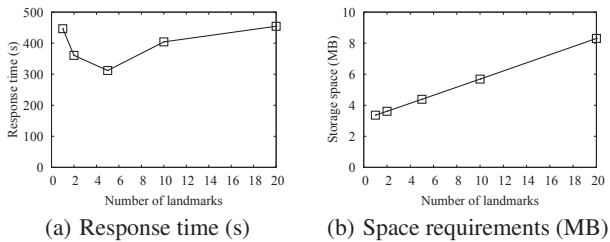

(a) Response time (s)  (b) Space requirements (MB)

**Figure 5: LM fine-tuning (Argentina)**

## 7.3 Comparison on Real Datasets

We first compare AF, LM, CI and PI on the Argentina network, and present the results in Table 3. The response time is dominated by the PIR cost, while the communication time and the client-side computations account only for a small fraction of the total time. With CI the client receives query results 3 times faster than AF and LM. PI is even more efficient, outperforming the baseline schemes by more than 5 times. To interpret performance, the table also shows the number of PIR accesses in the region data and network index files. AF and LM require reading more than half the database for each query in order to abide by the query plan. Turning to our methods, CI incurs 5 times more PIR accesses than PI; however, they have a smaller difference in response time. This is because PI performs retrievals on a much larger network index file than CI, i.e., each access costs more. Table 3 also presents the space requirements of the schemes. PI has the largest database, due to its voluminous index.

Figure 6 illustrates the response time of an obfuscation method (OBF) based on [22]. To reduce the amount of information leaked, we form the obfuscation sets of $s$ and $t$ with decoys randomly and uniformly chosen in the road network, instead of selecting locations near them (as in [22]). The figure shows the overall response time versus the size of obfuscation sets $S$ and $T$ (where $|S| = |T|$) on the Argentina network. For $|S|$ and $|T|$ in the order of tens, OBF is less efficient than our schemes (CI and PI, represented by horizontal lines) due to its large communication and server processing costs. OBF provides weak privacy (there are $|S|$ and $|T|$ specific positions where $s$ and $t$ could lie). It is therefore not directly comparable to the strongly secure methods we study and is not considered further.

**Table 3: Components of response time (Argentina)**

| Method | AF | LM | CI | PI |
|---|---|---|---|---|
| Response time (s) | 324.18 | 311.93 | 105.45 | 58.17 |
| PIR time (s) | 272.56 | 265.38 | 88.09 | 54.21 |
| Communication time (s) | 51.47 | 46.43 | 17.34 | 3.94 |
| Client-side computations (s) | 0.12 | 0.02 | 0.02 | 0.01 |
| PIR page accesses of the region data file | 595 of 820 | 536 of 1,096 | 193 of 775 | 2 of 775 |
| PIR page accesses of the network index file | 0 | 0 | 2 of 1,327 | 36 of 274,788 |
| Total storage space (MB) | 3.28 | 4.38 | 8.40 | 1,102 |

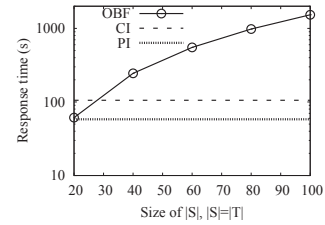

**Figure 6: Effect of $|S|$ on OBF, with $|S| = |T|$ (Argentina)**

Figure 7 compares AF, LM, CI, and PI on different road networks (namely, Old., Ger., Arg.). The results fall in line with those in Table 3 and establish the superiority of PI. However, its storage space grows rapidly with the size of the road network. For larger datasets (Den., Ind., Nor.), the network index file of PI exceeds the 2.5 GB limit, rendering the scheme inapplicable, and calling for its tunable-size variants, HY and PI*. We postpone the investigation of this larger-network case to Section 7.5.

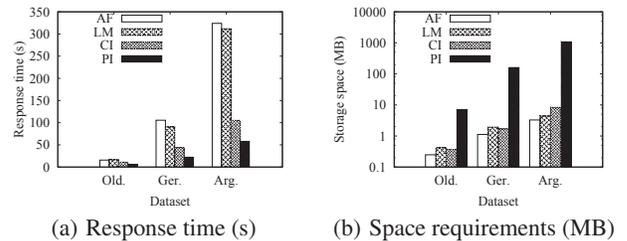

(a) Response time (s)  (b) Space requirements (MB)

**Figure 7: Performance on different road networks**

## 7.4 Effectiveness of Optimizations

In this section we examine the effectiveness of the two enhancements introduced in Section 5, namely packed partitioning (Section 5.6) and index compression (Section 5.5).

In Figure 8 we compare versions of CI and PI with packed versus plain KD-tree partitioning; the plain KD-tree variants are denoted as CI-P and PI-P. Figure 8(a) illustrates space utilization in the region data file $F_d$ for different networks. Packing achieves over 95% utilization in $F_d$ in all cases. In contrast, the utilization for CI-P and PI-P can be as low as 51% (for Ger.). The higher utilization in CI and PI implies that $F_d$ contains a smaller number of regions (equivalently, takes up fewer pages), leading also to a smaller network index file $F_i$. This explains the vast reduction in database size for CI and PI in Figure 8(c) (note the logarithmic scale). Importantly, the packed partitioning improves significantly the response time of CI. This is because most accesses in CI are performed in the region data file, which is significantly smaller than CI-P. On the other hand, the response time of PI is relatively unaffected, because it reads only two pages from $F_d$.

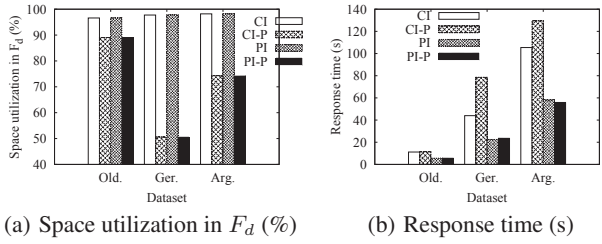

(a) Space utilization in $F_d$ (%)   (b) Response time (s)

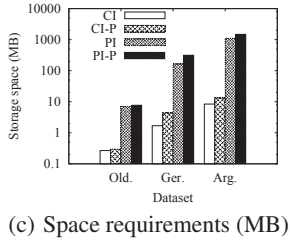

(c) Space requirements (MB)

**Figure 8: Effect of packed partitioning on CI and PI**

In Figure 9 we compare the full-fledged CI and PI against their counterparts without index compression (denoted as CI-C, PI-C). Figure 9(b) shows that the compression reduces storage space significantly, especially for datasets with larger sizes. The PI-C scheme (without compression) produces a network index file beyond 2.5 GByte on Argentina network. Its bar is marked as 'Nil', indicating that it is inapplicable in this case – on the other hand, PI (with compression) applies successfully. Figure 9(a) illustrates that index compression improves the response time of PI but not that of CI. The reason is explained by the number of PIR page accesses in Table 3, namely, that most accesses in CI are made on the region data file (whose size remains unchanged), whereas most accesses in PI are made on the network index file (whose size is reduced).

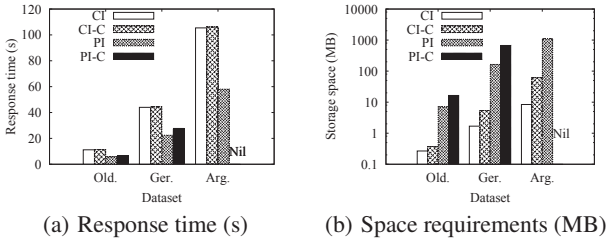

(a) Response time (s)   (b) Space requirements (MB)

**Figure 9: Effect of compression on CI and PI**

## 7.5 Experiments on Larger Networks

In this section we consider larger road networks (Den., Ind., Nor.), where PI is inapplicable due to the size of its network index file. The place of PI is taken by Hybrid Scheme (HY) and Clustered PI (PI*). Both HY and PI* enable tuning the network index size. We investigate the space-time trade-off in each of them before comparing with alternatives.

We first justify the design rationale of HY by examining the cardinalities $|S_{i,j}|$ of region sets in the network index of CI. In Figure 10(a) we run CI on Denmark and plot a histogram showing the number of $S_{i,j}$ sets for different cardinalities. Since the largest region set has cardinality 229 (i.e., $m = 229$), the query plan of CI requires fetching 229+2 disk pages from the region data file, leading to a long response time (see horizontal line in Figure 10(b)).

However, the majority of region sets have cardinalities below 100, implying that there is plenty of space for improvement with HY.

In Figures 10(b) and 10(c) we use a cardinality threshold as the tuning parameter for HY – any region set $S_{i,j}$ with cardinality greater than the threshold is replaced by the corresponding $G_{i,j}$ subgraph. The threshold essentially controls the number of PIR retrievals in the query plan. The smaller this parameter, the shorter the response time (and the larger the space required). In terms of response time, the best threshold value is the smallest for which the network index file does not exceed the maximum size supported. Although HY is meant to be a compromise between the time-intensive CI and the space-intensive PI, for very large threshold values it becomes slower than CI. The reason is that HY stores both region data and network index into the same file (of increased size), which means that each PIR access in HY is more expensive than in CI.

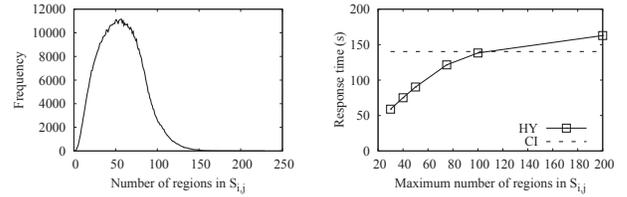

(a) Distribution of $|S_{i,j}|$ in CI   (b) Response time (s)

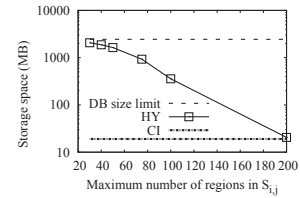

(c) Space requirements (MB)

**Figure 10: Performance of HY vs. limit on $|S_{i,j}|$ (Denmark)**

The performance of PI* is determined by the number of cluster pages, i.e., the number of disk pages allocated per region in the region data file. Figure 11 shows the response time and space requirements of PI* on Denmark network. As explained at the end of Section 6, when the cluster size increases, the response time rises (and the space consumption drops). PI* achieves its best performance at the smallest cluster size (2) for which the network index file is within the size limit of the PIR interface.

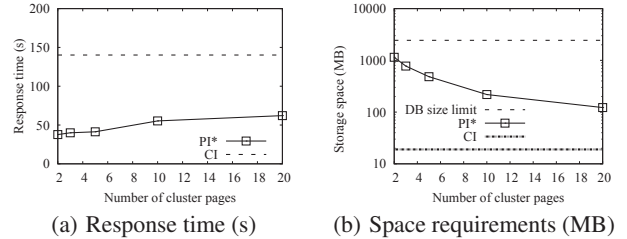

(a) Response time (s)   (b) Space requirements (MB)

**Figure 11: Performance of PI* vs. cluster size (Denmark)**

In the last experiment we compare CI, HY, and PI* on three large networks. As AF and LM are particularly inefficient, we exclude them from the charts. We tune both HY and PI* for the best running time, while space requirements stay within the PIR-imposed bound. The results in Figure 12 show that PI* achieves the fastest query processing in all cases.

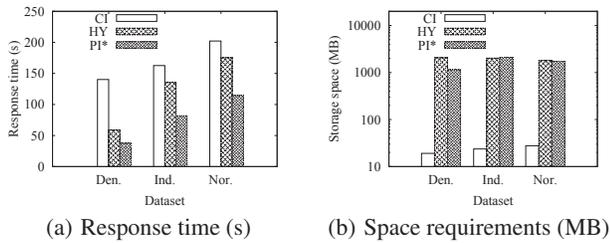

(a) Response time (s)  (b) Space requirements (MB)

**Figure 12: Performance on larger networks**

## 8. CONCLUSIONS

In this paper we propose the first PIR-based framework for private shortest path computation. Our objective is that the processing server answers client queries without inferring any information about them. We propose a suite of schemes that are readily deployable (as they rely on off-the-shelf PIR building blocks) and empirically evaluate them on large, real road networks. While performance is reasonable for the degree of privacy achieved, the space and time overheads imposed are significantly higher than unprotected query processing. A challenging topic for future work is to reduce these overheads. A possible direction is via (lossless or lossy) compression of network data, taking into account their characteristics/structure. Another is the development of approximate schemes with bounded cost deviation from the actual shortest path.